\documentclass[aps, prd, twocolumn, lengthcheck, superscriptaddress, showpacs, letterpaper, nofootinbib]{revtex4-1}

\usepackage{epsfig}
\usepackage[usenames]{color}
\usepackage{graphicx}
\usepackage{amsmath}
\usepackage{epstopdf}

\allowdisplaybreaks

\begin{document}

\title{Derivation of Brown-Rho scaling from scale-chiral perturbation theory}

\author{Yan-Ling Li}
\affiliation{Center for Theoretical Physics and College of Physics, Jilin University, Changchun,
130012, China}

\author{Yong-Liang Ma}
\email{yongliangma@jlu.edu.cn}
\affiliation{Center for Theoretical Physics and College of Physics, Jilin University, Changchun,
130012, China}

\date{\today}

\begin{abstract}
	
The medium modified hadron properties are studied by using the scale-chiral perturbation theory $\chi$PT$_\sigma$ in which the lightest scalar meson $f_0(500)$ is included as an explicit degree of freedom by regarding it as the Nambu-Goldstone boson of scale symmetry both spontaneously broken and  explicitly broken by the QCD trace anomaly. We derive Brown-Rho scaling at the leading order of scale symmetry from $\chi$PT$_\sigma$ and formulate how to make higher-order scale-chiral corrections going beyond the leading order of scale symmetry. By taking into account the intrinsic density dependence given by the dilaton condensate in the ``bare" parameters of the Lagrangian, the medium modified hadron properties are investigated. Relying on available experimental information and certain reasonable assumptions, we arrive at  the leading order scale symmetric effective theory that can be confronted with nature .
\end{abstract}

\maketitle

\section{Introduction}
\label{sec:Intro}

The phase structure of QCD at extreme conditions such as high density and/or temperature is still a question puzzling us. At low energy of QCD, the hadron physics can be systematically studied by using the effective theories based on the breaking of certain symmetries of QCD, such as the chiral symmetry and scale symmetry~\cite{MaRhoBook}.

Recently, assuming that there exists a nonperturbative infrared fixed point in QCD, a scale-chiral effective theory was proposed to include the lightest scalar meson $f_0(500)$ as the Nambu-Goldstone boson of the spontaneous breaking of scale symmetry of QCD (CT approach)~\cite{Crewther:2013vea}. After that, this approach was extended to include the vector mesons~\cite{Li:2016uzn} through the hidden local symmetry (HLS) approach~\cite{Bando:1984ej,Bando:1987br,Harada:2003jx}. Using this hidden local symmetrized scale-chiral effective theory, it was found that the baryonic matter from the skyrmion crystal approach accommodates scale restoration~\cite{Ma:2016nki}.

In nuclear matter,  the hadron properties are expected to be modified by medium. Given the scale-chiral effective theory $\chi$PT$_\sigma$ should be applicable in medium, the density effect could be intrinsically encoded -- inherited from QCD -- in ``bare" low energy constants of the effective theory, that is, endowed with intrinsic density dependence (IDD), to be distinguished from the density dependence coming from  mundane nuclear interactions.  We extend the notion of  Brown-Rho (BR) scaling~ \cite{Brown:1991kk} to a generalized framework in which the ``bare" parameters of effective field theory Lagrangian are endowed with the IDD inherited from QCD via  the vacuum expectation value of dilaton modified by density. It should be noted that when one confronts the BR scaling with experimental measurements, the matching should be done in the sense of correlation functions and all the medium modifications including including the quantum corrections such as higher order nuclear correlations should be taken into account  in the consistent way. So far this scaling has not been  -- and cannot -- be ruled out by the experiments thus far far analyzed~\cite{Brown:2005ka,Brown:2005kb,Brown:2009az}.

In the CT approach, the trace anomaly effect is included in both the leading current algebra terms and the dilaton potential terms. This formulation introduces  far too large a  number of  parameters in the scale-chiral effective theory to render the systematical application of this theory feasible. Alternatively to the CT approach, as suggested by Yamawaki~\cite{Yamawaki:2016qux}, hidden scale symmetry could be revealed in linear sigma model with the dilaton ``emerging" in a conformal compensator field with the scale-symmetry explicit breaking effect entirely included in the dilaton potential term. This approach makes  the dilaton-limit fixed-point at which the linear realization of chiral symmetry is restored  accessible by tweaking external conditions, e.g., high density~\cite{Sasaki:2011ff,Paeng:2011hy,Paeng:2015noa}. We shall show that this approach arises at the leading order of the scale symmetry (LOSS) in the CT scheme.

During the past decades, the medium modified hadron properties have been investigated experimentally. Therefore, in combination with the past progress and the IDD from the present scale-chiral effective theory~\cite{Li:2016uzn}, we expect to get some insights into the information of the low energy constants, i.e., ``bare parameters" of the effective Lagrangian as well as the basic nonperturbative properties of QCD at low density, such as $\lesssim n_{1/2} \simeq 2 n_0$ with $n_0$ being the normal nuclear density and $n_{1/2}$ being the density at which the topological object skyrmion in the theory factorizes into two half-skyrmions and after that, it is difficult to expect the validity of the present conclusion due to the topological change (for a recent review, see, e.g., Ref.~\cite{Ma:2016gdd}). And this would render feasible a more reliable study of the QCD phase structure. This is the main purpose of our present work.

What we find in the present work is that, in combination with the experimental observation and basic requirement from QCD, the general scale-chiral effective theory can be reduced to the LOSS and the BR scaling of the hadron properties in medium can be accessed in the regime where the  mean field approximation can be valid~\footnote{ There has been quite a bit of confusion on the meaning of ``BR scaling" in the literature. Since it will figure principally in this paper, we give a brief clarification of what its definition is. As clearly stated in \cite{Brown:1991kk}, it is the density-scaling behavior of the ``bare parameters" of the effective field theory (EFT) Lagrangian with which quantum calculations are to be performed. The properties of the parameters are inherited from QCD by the matching of the correlators of the EFT and QCD at a matching scale in the matter-free vacuum and the density dependence of the parameters reflects the intrinsic behavior of the QCD condensates as the vacuum is modified by the density in nuclear interactions.   As such, those parameters are {\it not} physical quantities and the scaling property, in general, cannot be extracted directly from experimental data since the latter subsumes all orders of those parameters resulting from suitable quantum calculations. Now given the $bs$HLS Lagrangian as constructed in \cite{Li:2016uzn}, with the intrinsic density dependence suitably incorporated, doing mean-field calculation is found to be tantamount to doing Landau Fermi-liquid theory (\`a la single-decimation Wilsonian renormalization group). When this approach is valid, then the experimental information can be related to the tree-order quantities in $bs$HLS. It turns out that this approximation works fairly well at relatively low density in the vicinity of nuclear matter density as is familiar in nuclear community in the name of ``energy density functional theory."  It however cannot be valid at high density as one approaches phase transitions as is discussed in \cite{MaRhoBook,PKLMR}.}.   Therefore, it is safe to apply the LOSS to study the medium modified hadron properties at this moment. It should be noted that the reduction to LOSS and derivation of BR scaling are performed in the even-parity part of the scale-chiral effective theory. For the odd-parity part of the effective theory such as the homogeneous Wess-Zumino part which encodes the omega meson effect, one should incorporate the correction to the LOSS arising form the explicit scale symmetry breaking to realize the partial restoration of chiral and scale symmetry in nuclear matter~\cite{Park:2008zg,Ma:2016nki}.

This paper is organized as follows: In Sec.~\ref{sec:Mmhp}, we write down the scale-chiral effective theory which will be used in the present work. We derive the general medium modified hadron properties, such as hadron masses, decay constants, coupling strengths, in terms of IDD of the LECs of the scale-chiral effective theory in Sec.~\ref{sec:mediumH}. We discuss the implications of the LECs with respect to both the theoretical calculations and experimental observations in Sec.~\ref{sec:impli}. Our summary and discussion are given in Sec.~\ref{sec:dis}.


\section{Scale-chiral effective theory}
\label{sec:Mmhp}

For deriving the IDD of the hadron properties from the scale-chiral effective theory, we first construct the scale-hidden local symmetry, HLS$_\sigma$ for mesons, and baryonic-scale-hidden local symmetry $bs$HLS including baryons. The HLS$_\sigma$ is constructed by coupling the dilaton arising from the spontaneous breaking of scale symmetry to the HLS. In HLS, one decomposes the pion field $U(x) = \exp(i\Pi/f_\pi)$ as~\cite{Harada:2003jx}
\begin{eqnarray}
U(x) & = & \xi_L^\dagger \xi_R
\end{eqnarray}
where $\Pi = \pi^a\lambda^a$ with $\lambda^a$ being the Gell-Mann matrix. Under chiral transformation, they transform as
\begin{eqnarray}
U(x) & \to & g_L U(x) g_R^\dagger, \nonumber\\
\xi_{L,R}(x) & \to & h(x)\xi_{L,R}(x) g_{L,R}^\dagger
\end{eqnarray}
with $g_{L,R} \in SU(3)_{L,R}$ and $h(x) \in U(3)_{\rm local}$ been used in the present work. The HLS Lagrangian is constructed by the following two Mauer-Cartan  $1$-forms
\begin{eqnarray}
\hat{\alpha}_{\parallel\mu} & = & \frac{1}{2i}(\mathcal{D}_\mu \xi_R \cdot
\xi_R^\dag +
\mathcal{D}_\mu \xi_L \cdot \xi_L^\dag), \nonumber\\
\hat{\alpha}_{\perp\mu} & = & \frac{1}{2i}(\mathcal{D}_\mu \xi_R \cdot
\xi_R^\dag - \mathcal{D}_\mu \xi_L \cdot \xi_L^\dag)
\ , \label{eq:1form}
\end{eqnarray}
where the covariant derivatives are defined as
\begin{eqnarray}
\mathcal{D}_\mu \xi_{L} & = & \partial_\mu \xi_{L} - i V_\mu \xi_{L} + i \xi_{L}\mathcal{L}_\mu ,\nonumber\\
\mathcal{D}_\mu \xi_{R} & = & \partial_\mu \xi_{R} - i V_\mu \xi_{R} + i \xi_{R}\mathcal{R}_\mu .
\end{eqnarray}
The external gauge fields $\mathcal{L}_{\mu}$ and $\mathcal{R}_{\mu}$ can be expressed in terms of $W_{\mu}$,$ Z_{\mu}$ and $ A_{\mu}$ (photon) through
\begin{eqnarray}
\mathcal{L}_{\mu} & = & e Q A_{\mu} +\frac{g_{2}}{\cos\theta_{W}}(T_{z}-\sin^{2}{\theta_{W}})Z_{\mu}\nonumber\\
& &{} +\frac{g_{2}}{\sqrt{2}}\left( W_{\mu}^{+}T_{+} + W_{\mu}^{-}T_{-}\right),\nonumber\\
\mathcal{R}_{\mu} & = & e Q A_{\mu}-\frac{g_{2}}{\cos\theta_{W}}\sin^{2}{\theta_{W}}Z_{\mu},
\end{eqnarray}
where $ e $,$g_{2}$ and $ \theta_{W}$ are the electromagnetic coupling constant, the weak gauge coupling constant and the weak mixing angle, respectively. The electric charge matrix $Q$ is given by $Q = \frac{1}{3} {\rm diag}( 2, -1, -1 )$, $T_{z} = \frac{1}{2} {\rm diag}( 1, -1, -1 )$ and the nonzero matrix elements of $T_{+}$ are $(T_{+})_{12} = V_{ud},  (T_{+})_{13} = V_{us}$. After gauge-fixing, in the unitary gauge with
\begin{eqnarray}
\xi_L^\dag & = & \xi_R \equiv \xi = e^{i\pi/(2f_\pi)},
\label{eq:hiddenbreaking}
\end{eqnarray}
the HLS gauge field can be identified with the massive vector mesons in the following matrix form
\begin{eqnarray}
V_\mu & = & \frac{g}{\sqrt{2}}\left(
                                \begin{array}{ccc}
                                  \frac{1}{\sqrt{2}}(\rho_\mu^0 + \omega_\mu) & \rho_\mu^+ & K_\mu^{\ast,+} \\
                                  \rho_\mu^- & {}-\frac{1}{\sqrt{2}}(\rho_\mu^0 - \omega_\mu) & K_\mu^{\ast,0} \\
                                  K_\mu^{\ast,-} & \bar{K}_\mu^{\ast,0} & \phi_\mu \\
                                \end{array}
                              \right)
.\label{eq:vectormeson}
\end{eqnarray}
By using the 1-forms~\eqref{eq:1form} and the energy momentum tensor of the vector meson field $V_{\mu\nu} = \partial_\mu V_\nu - \partial_\nu V_\mu = i [V_\mu,V_\nu]$, one can easily construct the HLS Lagrangian~\cite{Bando:1984ej,Bando:1987br,Harada:2003jx}.

To construct the HLS$_\sigma$, we introducing the conformal compensator $\chi$ which is a chiral singlet and transforms under scaling $x_\mu \to \lambda^{-1} x_\mu$ as
\begin{eqnarray}
\chi(x) & \to & \lambda \chi(\lambda^{-1}x).
\end{eqnarray}
In terms of the NGB $\sigma$ associated with the spontaneous breaking of scale symmetry, $\chi$ field is written as
\begin{eqnarray}
\chi(x) = f_\sigma e^{\sigma/f_\sigma},
\end{eqnarray}
where $f_\sigma$ is the $\sigma$ decay constant. The ${\rm HLS}_{\sigma}$ is written down by coupling the $\chi$ field to the HLS Lagrangian.

With respect to the scale symmetry, scale symmetry breaking and explicit chiral symmetry, the leading order ${\rm HLS}_{\sigma}$ Lagrangian which will be used in the present work is written as~\cite{Li:2016uzn}:
\begin{eqnarray}
\cal L_{{\rm HLS}_\sigma}^{\rm LO}& = & {\cal L}_{{\rm HLS}_\sigma;{\rm inv}}^{d=4} + {\cal L}_{{\rm HLS}_\sigma;{\rm anom}}^{d > 4} + {\cal L}_{{\rm HLS}_\sigma;{\rm mass}}^{d < 4},\label{eq:CTHLS}
\end{eqnarray}
where ${\cal L}_{{\rm HLS}_\sigma;{\rm inv}}^{d=4}$ is the scale invariant part, ${\cal L}_{{\rm HLS}_\sigma;{\rm anom}}^{d > 4}$ is the part accounting for the trace anomaly in chiral limit and ${\cal L}_{{\rm HLS}_\sigma;{\rm mass}}^{d < 4}$ is the scale symmetry breaking part arising from the current quark mass. Explicitly, these three parts have expressions
\begin{widetext}
\begin{subequations}
\begin{eqnarray}
{\cal L}_{{\rm HLS}_\sigma; {\rm inv}}^{d=4} & = & f_\pi^2h_1 \left(\frac{\chi}{f_\sigma}\right)^2{\rm
Tr}[\hat{a}_{\perp\mu}\hat{a}_{\perp}^{\mu}] + a f_\pi^2 h_2  \left(\frac{\chi}{f_\sigma}\right)^2{\rm
Tr}[\hat{a}_{\parallel\mu}\hat{a}_{\parallel}^{\mu}] -
\frac{1}{2g^2}h_3{\rm Tr}[V_{\mu\nu}V^{\mu\nu}] \nonumber\\
& &{} + \frac{1}{2}h_4 \partial_\mu \chi \partial^\mu \chi + h_5 \left(\frac{\chi}{f_\sigma}\right)^4, \label{eq:CTHLS40}\\
{\cal L}_{{\rm HLS}_\sigma; {\rm anom}}^{d > 4} & = & f_\pi^2(1-h_1) \left(\frac{\chi}{f_\sigma}\right)^{2+\beta^\prime}{\rm
Tr}[\hat{a}_{\perp\mu}\hat{a}_{\perp}^{\mu}] + (1-h_2) f_\sigma^2\left(\frac{\chi}{f_\sigma}\right)^{2+\beta^\prime}{\rm
Tr}[\hat{a}_{\parallel\mu}\hat{a}_{\parallel}^{\mu}]\nonumber\\
& &{}  -
\frac{1}{2g^2}(1-h_3)\left(\frac{\chi}{f_\sigma}\right)^{\beta^\prime}{\rm Tr}[V_{\mu\nu}V^{\mu\nu}] + \frac{1}{2}(1-h_4) \left(\frac{\chi}{f_\sigma}\right)^{\beta^\prime} \partial_\mu \chi \partial^\mu \chi + h_6 \left(\frac{\chi}{f_\sigma}\right)^{4+\beta^\prime}, \label{eq:CTHLSg40}\\
{\cal L}_{{\rm HLS}_\sigma; {\rm mass}}^{d < 4} & = &{} \frac{f_\pi^2}{4} \left(\frac{\chi}{f_\sigma}\right)^{3-\gamma_m}{\rm Tr}\left(\hat{\mathcal{M}} + \hat{\mathcal{M}}^\dagger\right),\label{eq:CTHLSm40}
\end{eqnarray}
\end{subequations}
\end{widetext}
where $\hat{\mathcal{M}} \equiv \xi_L \mathcal{M} \xi_R^\dagger$ transforms under HLS as $\hat{\mathcal{M}} \to h(x) \hat{\mathcal{M}} h^\dagger(x)$ and $\mathcal{M} = {\rm diag}(m_\pi^2, m_\pi^2, 2m_K^2-m_\pi^2)$. Here $h$'s are unknown constants and $\beta^\prime$ is the anomalous dimension of the gluon energy momentum tensor squared, that signals scale symmetry explicit breaking~\cite{Crewther:2013vea}. Expanding the 1-forms \eqref{eq:1form} in terms of meson fields, one has
\begin{eqnarray}
\widehat{\alpha}_{\perp\mu} & = &
\frac{1}{f_\pi} \partial_\mu \pi
+ {\cal A}_\mu
- \frac{i}{f_\pi} \left[ {\cal V}_\mu \,,\, \pi \right]
+ \cdots
\ ,
\label{al perp exp}
\\
\widehat{\alpha}_{\parallel\mu} & = &
{} - V_\mu
+ {\cal V}_\mu
- \frac{i}{2f_\pi^2} \bigl[ \partial_\mu \pi \,,\, \pi \bigr]
- \frac{i}{f_\pi} \left[ {\cal A}_\mu \,,\, \pi \right]
+ \cdots
\ ,
\nonumber\\
\label{al para exp}
\end{eqnarray}
where
${\cal V}_\mu = \left({\cal R}_\mu + {\cal L}_\mu\right)/2$ and ${\cal A}_\mu = \left({\cal R}_\mu - {\cal L}_\mu\right)/2$. So that, the hadron properties such as mass, interactions and decay widths can be written down explicitly.

The above construction of the HLS$_\sigma$ can be easily extended to include the baryon octet
\begin{eqnarray}
B & = & \left( \begin{array}{ccc} \frac{1}{\sqrt{2}} \Sigma^0 + \frac{1}{\sqrt{6}} \Lambda & \Sigma^+                                                  & p \\
		\Sigma^-                                                 & -\frac{1}{\sqrt{2}} \Sigma^0 + \frac{1}{\sqrt{6}} \Lambda & n \\
		\Xi^-                                                    & \Xi^0                                                     & -\frac{2}{\sqrt{6}} \Lambda  \end{array} \right).
\end{eqnarray}
And, upto the scale-chiral order $O(p)$, the scale-hidden local symmetric theory involving baryons, $bs$HLS, is written as~\cite{Li:2016uzn}
\begin{eqnarray}
{\cal L}_{bs{\rm HLS}} & = & {\cal L}_{bs{\rm HLS}; {\rm inv}}^{d = 4} + {\cal L}_{bs{\rm HLS}; {\rm anom}}^{d > 4},
\label{eq:LagbsHLS}
\end{eqnarray}
with
\begin{widetext}
\begin{subequations}
\begin{eqnarray}
{\cal L}_{bs{\rm HLS}; {\rm inv}}^{{\rm LO}; d = 4} & = & g_1{\rm Tr}\left( \bar{B} i \gamma_\mu D^\mu B\right) - \tilde{\mathring{m}}_B \frac{\chi}{f_\sigma}{\rm Tr}\left( \bar{B} B\right) - \tilde{g}_{A_1}{\rm Tr}\left( \bar{B} \gamma_\mu \gamma_5 \left\{ \hat{\alpha}_\perp^\mu,  B \right\} \right) \nonumber\\
& &{} - \tilde{g}_{A_2}{\rm Tr}\left( \bar{B} \gamma_\mu \gamma_5 \left[ \hat{\alpha}_\perp^\mu,  B \right] \right) + \tilde{g}_{V_1}{\rm Tr}\left( \bar{B} \gamma_\mu \left\{ \hat{\alpha}_\parallel^\mu,  B \right\} \right) + \tilde{g}_{V_2}{\rm Tr}\left( \bar{B} \gamma_\mu \left[ \hat{\alpha}_\parallel^\mu,  B \right] \right) \, ,\label{LBCTd4} \\
{\cal L}_{bs{\rm HLS}; {\rm anom}}^{{\rm LO}; d > 4} & = & \Big[(1-g_1){\rm Tr}\left( \bar{B} i \gamma_\mu D^\mu B\right) - (\mathring{m}_B - \tilde{\mathring{m}}_B) \frac{\chi}{f_\sigma} {\rm Tr}\left( \bar{B} B\right) - (g_{A_1} - \tilde{g}_{A_1}){\rm Tr}\left( \bar{B} \gamma_\mu \gamma_5 \left\{ \hat{\alpha}_\perp^\mu,  B \right\} \right)\nonumber\\
& &\;{} - (g_{A_2} - \tilde{g}_{A_2}){\rm Tr}\left( \bar{B} \gamma_\mu \gamma_5 \left[ \hat{\alpha}_\perp^\mu,  B \right] \right) + (g_{V_1} - \tilde{g}_{V_1}){\rm Tr}\left( \bar{B} \gamma_\mu \left\{ \hat{\alpha}_\parallel^\mu,  B \right\} \right) \nonumber\\
& &\;{} + (g_{V_2} - \tilde{g}_{V_2}){\rm Tr}\left( \bar{B} \gamma_\mu \left[ \hat{\alpha}_\parallel^\mu,  B \right] \right) \Big]\left(\frac{\chi}{f_\sigma}\right)^{\beta^\prime} ,
\label{LBCT2}
\end{eqnarray}
\end{subequations}
\end{widetext}
where
\begin{equation}
D^\mu B = \partial^\mu B -i \left[ V^\mu, B \right]\, .
\end{equation}
with $V_\mu$ being the vector meson field~\eqref{eq:vectormeson}.

In the general Lagrangian, except the parameters in HLS, there are nine parameters $f_\sigma, \beta^\prime , \gamma_m$ and $h_i (i = 1, \cdots 6)$. The $\sigma$ decay constant $f_\sigma$ is taken to be at the same order as $f_\pi$. The minimization of the potential, the $\sigma$ meson mass and the $\sigma \to \pi\pi$ decay can reduce the number of the free parameters to five, $\beta^\prime, \gamma_m, h_2, h_3$ and $h_4$. In the baryonic sector, there are parameters  $\mathring{m}_B, \tilde{\mathring{m}}_B, g_1, g_{A_1}, \tilde{g}_{A_1}, g_{A_2}, \tilde{g}_{A_2}, g_{V_1}, \tilde{g}_{V_1}, g_{V_2}$ and $\tilde{g}_{V_2}$. The nucleon mass in chiral limit $\mathring{m}_B$ can be fixed from the calculation of nuclear matter~\cite{Li:2008hp} and $g_{A_1}$ and $g_{A_2}$ can be estimated by using the $B \to B^\prime + e^- + \bar{\nu}_e$~\cite{Borasoy:1998pe} so that there are nine parameters in the baryonic sector.

Generally, the Lagrangian for the matter part (pseudoscalar meson, vector meson and baryon) from the CT approach can be written as
\begin{equation}
{\cal L} = \left[\kappa + (1 - \kappa)\left(\frac{\chi}{f_\sigma}\right)^{\beta^\prime}\right]\bar{{\cal L}},
\end{equation}
where $\bar{{\cal L}}$ is the scale invariant Lagrangian constructed by multiplying the conformal compensator $\chi$ to the mass dimension $m$ (with $m \leq 4$) Lagrangian ${\cal L}^{(m)}$ through
\begin{equation}
\bar{{\cal L}} = \left(\frac{\chi}{f_\sigma}\right)^{4-m}{\cal L}^{(m)}.
\end{equation}
The hidden scale symmetry suggested in Ref.~\cite{Yamawaki:2016qux} can be accessed in two ways, the one is by setting $\beta^\prime \ll 1$ and the other one is by choosing $\kappa \approx 1$. From the skyrmion crystal simulation of the nuclear matter, it was found that the former choice is not acceptable~\cite{Ma:2016nki}. We will investigate the viability of the latter choice which gives rise to the LOSS with respect to the present measurement and other theoretical progress.

\section{Medium modified hadron properties}

\label{sec:mediumH}

When a hadron is put into a medium such as the nuclear matter, its properties are expected to be modified due to the interaction between the hadron and medium and consequently depend on the medium density. Given the effective theory should be applicable in medium, the density effect should be embedded in the low energy constants of the effective theory, that is, become IDD quantities. As mentioned above, one can infer this IDD entirely  from the vacuum expectation value of the dilaton field, $\langle\chi\rangle^\ast = f_\sigma^\ast$  at low density at which the hadron dynamics is still applicable. Through out this work the density dependence is indicated by the asterisk and the symbols without asterisk denote the value at the matter free space. In this work, we expand $\chi$ with respect to its expectation value in medium $\langle \chi \rangle^\ast$ to study the medium modified quantities.

\subsection{Hadron properties in medium}


We first define the physical hadron fields in medium. For this purpose, we consider the possible terms in the Lagrangian~\eqref{eq:CTHLS} and \eqref{eq:LagbsHLS} contributing to the hadron kinetic terms with expression
\begin{eqnarray}
{\cal L}_{\rm kin}& = & \left[h_1+ (1 - h_1)\left(\frac{\langle \chi \rangle^\ast}{f_\sigma}\right)^{\beta^{'}}\right]\left(\frac{\langle \chi \rangle^\ast}{f_\sigma}\right)^{2} {\rm Tr}\left( \partial_\mu \Pi \partial^\mu \Pi \right) \nonumber\\
& &{} + \frac{1}{2}\left[ h_4+ (1 - h_4)\left(\frac{\langle \chi \rangle^\ast}{f_\sigma}\right)^{\beta^{'}}\right]\left(\frac{\langle \chi \rangle^\ast}{f_\sigma}\right)^{2} \partial_\mu \sigma \partial^\mu \sigma \nonumber\\
& &{} -\frac{1}{2g^2}\left[h_3+(1-h_3)\left(\frac{\langle \chi \rangle^\ast}{f_\sigma}\right)^{\beta^\prime}\right]{\rm Tr}\left[V_{\mu\nu}V^{\mu\nu}\right] \nonumber\\
& &{} + \left[g_1+ (1 - g_1)\left(\frac{\langle \chi \rangle^\ast}{f_\sigma}\right)^{\beta^{'}}\right] {\rm Tr}\left( \bar{B} i \gamma_\mu \partial^\mu B\right).
\end{eqnarray}
Therefore, the physical hadron fields in-medium can be defined as
\begin{eqnarray}
\tilde{\Pi} = {Z}_{3 \pi}\Pi ,\;\; \tilde{\sigma} ={Z}_{3 \sigma}\sigma ,\;\; \tilde{\rho}_\mu ={Z}_{3 \rho}\rho_\mu, \;\; \widetilde{B} = {Z}_{3B}B ,
\end{eqnarray}
where the wave function normalization factors are
\begin{eqnarray}
{Z}_{3\pi}^{2} & = & \left[h_1+ (1 - h_1)\left(\frac{\langle \chi \rangle^\ast}{f_\sigma}\right)^{\beta^{'}}\right]\left(\frac{\langle \chi \rangle^\ast}{f_\sigma}\right)^{2}  ,\label{eq:COF3pi} \nonumber\\
{Z}_{3 \sigma}^{2} & = & \left[ h_4+ (1 - h_4)\left(\frac{\langle \chi \rangle^\ast}{f_\sigma}\right)^{\beta^{'}}\right]\left(\frac{\langle \chi \rangle^\ast}{f_\sigma}\right)^{2},\label{eq:COF3si} \nonumber\\
{Z}_{3 \rho}^{2} & = & h_3+ (1 - h_3)\left(\frac{\langle \chi \rangle^\ast}{f_\sigma}\right)^{\beta^{'}}, \nonumber\\
{Z}_{3B}^{2} & = & g_1+ (1 - g_1)\left(\frac{\langle \chi \rangle^\ast}{f_\sigma}\right)^{\beta^{\prime}}.
\label{eq:COF3rho}
\end{eqnarray}
From the redefinition of the physical fields, we obtain the medium modified decay constants of sigma and pion mesons as
\begin{eqnarray}
f_\sigma^\ast = Z_{3 \sigma} f_\sigma , \quad  f_\pi^\ast = Z_{3 \pi} f_\pi.
\end{eqnarray}
In addition, from this normalization, one can define a new HLS gauge coupling constant through
\begin{eqnarray}
g^\ast & = &{} Z_{3\rho}^{-1}g,
\end{eqnarray}
such that energy-momentum tensor part of the HLS Lagrangian keeps intact.


We next derive the medium modified sigma meson mass. For this purpose, we first consider the relation between $h_5$ and $h_6$ in medium in the chiral limit. The minimization of the dilaton potential yields
\begin{eqnarray}
4h_5 + (4 + \beta^\prime)h_6 \left(\frac{\langle \chi \rangle^\ast}{f_\sigma}\right)^{\beta^\prime} & = & 0.
\end{eqnarray}
This indicates that, both or one of $h_5$ and $h_6$ should depend on density. To have a lower bounded potential, one can write
\begin{eqnarray}
h_5 & = & (4 + \beta^\prime)c \left(\frac{\langle \chi \rangle^\ast}{f_\sigma}\right)^{\beta^\prime}, \quad h_6 = {} -4c ,
\label{eq:relationc3c4}
\end{eqnarray}
with $c > 0$ to have a lower bound of the dilaton potential. Then, the medium modified sigma mass is obtained as
\begin{eqnarray}
m_\sigma^{\ast 2} & = & \left(\frac{\langle \chi \rangle^\ast}{f_\sigma}\right)^{4 + \beta^\prime}\frac{1}{Z_{3\sigma}^2}m_\sigma^2,
\label{eq:msigmas}
\end{eqnarray}
with $m_\sigma^2 = 4c\beta^\prime(4+\beta^\prime)/f_\sigma^2$ being the sigma meson mass in the matter free space.

For the pseudoscalar meson, we obtain the following expression contributing to the mass term
\begin{eqnarray}
{\cal L}_{\rm mass}^{\pi, K} & = & {} - \left(\frac{\langle \chi \rangle^\ast}{f_\sigma}\right)^{3 - \gamma_m}{\rm Tr}\Big(\mathcal{M}\Pi^2\Big).
\end{eqnarray}
So that, the medium modified pseudoscalar meson masses are
\begin{eqnarray}
m_\Pi^{\ast 2} & = & {} \left(\frac{\langle \chi \rangle^\ast}{f_\sigma}\right)^{3 - \gamma_m}m_\Pi^2 \frac{1}{Z_{3\pi}^2}.
\end{eqnarray}

For the vector meson mass, we have the contributing term
\begin{eqnarray}
{\cal L}_{\rm mass}^{V} & = &  \left[ h_2  + (1-h_2) \left(\frac{\langle\chi\rangle^\ast}{f_\sigma}\right)^{\beta^\prime}\right] \left(\frac{\langle\chi\rangle^\ast}{f_\sigma}\right)^2\nonumber\\
& &{} \times a f_\pi^2{\rm
	Tr}[V_{\mu}V^{\mu}].
\end{eqnarray}
Therefore the mass of the vector meson in medium is
\begin{eqnarray}
m_\rho^{\ast 2} & = & m_\rho^2 \left[ h_2 + (1-h_2) \left(\frac{\langle\chi\rangle^\ast}{f_\sigma}\right)^{\beta^\prime}\right]  \left(\frac{\langle\chi\rangle^\ast}{f_\sigma}\right)^2\frac{1}{Z_{3\rho}^2},
\nonumber\\
\end{eqnarray}
with $m_\rho^2 = a g^2 f_\pi^2$. Alternatively, one can also define the medium modified HLS parameter $a$ as follows
\begin{eqnarray}
a^{\ast} & = &a \left[ h_2 + (1-h_2) \left(\frac{\langle\chi\rangle^\ast}{f_\sigma}\right)^{\beta^\prime}\right]  \left(\frac{\langle\chi\rangle^\ast}{f_\sigma}\right)^2\frac{1}{Z_{3\pi}^2},
\end{eqnarray}
such that the medium modified vector meson mass can be expressed as
\begin{eqnarray}
m_\rho^{\ast 2} & = & a^{\ast}g^{\ast 2}f_\pi^{\ast 2}.
\end{eqnarray}

The baryon mass can be extracted from the Lagrangian
\begin{eqnarray}
{\cal L}_{B }^{\rm mass} & = & {} - \tilde{\mathring{m}}_B \frac{\langle \chi \rangle^\ast}{f_\sigma}{\rm Tr}\left( \bar{B} B\right) \nonumber\\
& &{} - (\mathring{m}_B - \tilde{\mathring{m}}_B) \left(\frac{\langle \chi \rangle^\ast}{f_\sigma}\right)^{1 + \beta^\prime} {\rm Tr}\left( \bar{B} B\right) ,
\end{eqnarray}
which yields the medium modified baryon mass as
\begin{eqnarray}
m_B^\ast & = & {} \left[\tilde{\mathring{m}}_B + (\mathring{m}_B - \tilde{\mathring{m}}_B) \left(\frac{\langle \chi \rangle^\ast}{f_\sigma}\right)^{ \beta^\prime} \right]\frac{\langle \chi \rangle^\ast}{f_\sigma}\frac{1}{{Z}_{3B}^{2}}.
\label{eq:massofb}
\end{eqnarray}


\subsection{Strong and electroweak couplings}

We next give the medium modified coupling constants. In terms of the physical fields and medium modified parameters defined above, the relevant Lagrangians for the strong interaction are expressed as
\begin{widetext}
\begin{eqnarray}
{\cal L}_{\sigma\pi\pi } & = & \left[2 h_1 + (2 + \beta^{\prime})(1 - h_1)\left(\frac{\langle \chi \rangle^\ast}{f_\sigma}\right)^{\beta^{\prime}} \right]\left(\frac{\langle \chi \rangle^\ast}{f_\sigma}\right)^{2}\frac{ 1}{f_\sigma^\ast}\frac{1}{Z_{3\pi}^2 }\tilde{\sigma} {\rm Tr}\left( \partial_\mu \tilde{\Pi} \partial^\mu \tilde{\Pi} \right) -\frac{(3-\gamma_m)}{ f_\sigma^\ast} \tilde{\sigma} {\rm Tr}\left(\mathcal{M}^{\ast}\tilde{\Pi}^2 \right), \\
{\cal L}_{\sigma\sigma\sigma} & = & \left[h_4 + (1 - h_4)\left(\frac{\langle \chi \rangle^\ast}{f_\sigma}\right)^{\beta^{\prime}}\left(1 + \frac{1}{2} \beta^{\prime} \right)\right]\left(\frac{\langle \chi \rangle^\ast}{f_\sigma}\right)^{2}\frac{1}{f_\sigma^\ast}\frac{1}{Z_{3\sigma}^2}\tilde{\sigma}\partial_\mu \tilde{\sigma} \partial^\mu \tilde{\sigma} \nonumber\\
& &{} + \frac{1}{6} \left[ 4c(4+\beta^\prime)\left(\frac{\langle \chi \rangle^\ast}{f_\sigma}\right)^{4+\beta^\prime}\left[16 - (4+\beta^\prime)^2\right] + \frac{f_{\pi}^{2}}{2}(3-\gamma_m)^3 (2m_K^{\ast 2} + m_\pi^{\ast 2})Z_{3\pi}^2 \right]\frac{1}{f_\sigma^{\ast 3}} \tilde{\sigma}^3 , \nonumber\\
{\cal L}_{\rho\pi\pi} & = & i a^\ast g^\ast{\rm Tr}[\tilde{\rho}_\mu \left[\partial^\mu \tilde{\pi}, \tilde{\pi}\right]],\nonumber\\
{\cal L}_{\sigma\rho\rho} & = &{} a g^{\ast 2}f_\pi^2\left[ 2 h_2 + (2+\beta^\prime)(1-h_2) \left(\frac{\langle\chi\rangle^\ast}{f_\sigma}\right)^{\beta^\prime} \right]\left(\frac{\langle\chi\rangle^\ast}{f_\sigma}\right)^2 \frac{1}{f_\sigma^\ast}\tilde{\sigma}{\rm Tr}[\tilde{\rho}_\mu \tilde{\rho}^\mu] \nonumber\\
& &{} -
\frac{1}{2}\beta^\prime(1-h_3)\left(\frac{\langle\chi\rangle^\ast}{f_\sigma}\right)^{\beta^\prime}\frac{1}{f_\sigma^\ast}\frac{1}{Z_{3\rho}^2}\tilde{\sigma}{\rm Tr}[(\partial_\mu \tilde{\rho}_\nu - \partial_\nu \tilde{\rho}_\mu)(\partial^\mu \tilde{\rho}^\nu - \partial^\nu \tilde{\rho}^\mu)] ,
\end{eqnarray}
\end{widetext}
where, for latter convenience, we keep the factor $\langle \chi \rangle^\ast/f_\sigma$.

Along the same method, one can derive the electroweak coupling. The straight forward expansion yields
\begin{eqnarray}
{\cal L}_{{\rm HLS}_\sigma} & = & \frac{g_2 f_\pi^\ast}{\sqrt{2}}\mbox{tr}\left[ \partial_\mu \tilde{\pi} \left( W^+_\mu T_+ + W^-_\mu T_- \right) \right] \nonumber\\
& &{} - 2ie\left[1 - \frac{a}{2} \frac{Z_{3\rho}^2}{Z_{3\pi}^2}\frac{m_\rho^{\ast 2}}{m_\rho^2}\right] \mbox{tr}\left[ \partial_\mu \tilde{\pi}  \left[ A_\mu Q \,,\, \tilde{\pi} \right] \right]\nonumber\\
& &{} - 2 e  \frac{m_\rho^{\ast 2}}{g^\ast} A^\mu \,\mbox{tr} \left[ \tilde{\rho}_\mu Q \right]  + \cdots \ .
\label{Lag exp}
\end{eqnarray}
which gives the medium modified electroweak couplings
\begin{eqnarray}
g_{\rho\gamma}^\ast & = & \left[h_2+ (1 - h_2)\left(\frac{\langle \chi \rangle^\ast}{f_\sigma}\right)^{\beta^{'}}\right]\left(\frac{\langle \chi \rangle^\ast}{f_\sigma}\right)^{2}\,\frac{  g_{\rho\gamma}}{Z_{3\rho}} = \frac{m_\rho^{\ast 2}}{g^\ast}\nonumber\\
g_{\gamma\pi\pi}^{\ast} & = & e \left[1 - \frac{a}{2} \frac{Z_{3\rho}^2}{Z_{3\pi}^2}\frac{m_\rho^{\ast 2}}{m_\rho^2}\right]\,\label{ggampp}\ .	
\end{eqnarray}

For the baryon-meson interactions, they can be easily derived as
\begin{widetext}
\begin{eqnarray}
\label{eq:SIGBB}
{\cal L}_{\sigma\bar{B} B} & = & - \tilde{\mathring{m}}_B \frac{\langle \chi \rangle^\ast}{f_\sigma}\frac{1}{f_\sigma^\ast}\frac{1}{Z_{3B}^2}\tilde{\sigma}{\rm Tr}\left( \bar{\tilde{B}} \tilde{B}\right) + (1-g_1) \left(\frac{\langle \chi \rangle^\ast}{f_\sigma}\right)^{\beta^{\prime}}\frac{\beta^{\prime} }{f_\sigma^\ast}\frac{1}{Z_{3B}^2}\tilde{\sigma}{\rm Tr}\left( \bar{\tilde{B}} i \gamma_\mu \partial^\mu \tilde{B}\right) \nonumber\\
& &{} - (\mathring{m}_B - \tilde{\mathring{m}}_B)  \left(\frac{\langle \chi \rangle^\ast}{f_\sigma}\right)^{1+\beta^{\prime}}\frac{1+\beta^\prime}{f_\sigma^\ast}\frac{1}{Z_{3B}^2} \tilde{\sigma}{\rm Tr}\left( \bar{\tilde{B}} \tilde{B} \right)\, , \nonumber\\
{\cal L}_{\pi\bar{B} B} & = &{} - \frac{1}{f_\pi^\ast}\left[\tilde{g}_{A_1} + (g_{A_1} - \tilde{g}_{A_1})\left(\frac{\langle\chi\rangle^\ast}{f_\sigma}\right)^{\beta^\prime}\right] \frac{1}{Z_{3B}^2}{\rm Tr}\left( \bar{\tilde{B}} \gamma_\mu \gamma_5 \left\{ \partial^\mu\tilde{\pi}, \tilde{B}\right\} \right) \nonumber\\
& &{} - \frac{1}{f_\pi^\ast}\left[ \tilde{g}_{A_2} + (g_{A_2} - \tilde{g}_{A_2})\left(\frac{\langle\chi\rangle^\ast}{f_\sigma}\right)^{\beta^\prime}\right] \frac{1}{Z_{3B}^2}{\rm Tr}\left( \bar{\tilde{B}} \gamma_\mu \gamma_5 \left[ \partial^\mu\tilde{\pi}, \tilde{B}\right] \right) ,\nonumber\\
{\cal L}_{V\bar{B} B} & = & \left\{\left(g_1 - \tilde{g}_{V_2}\right) + \left[(1-g_1) - (g_{V_2} - \tilde{g}_{V_2})\right]\left(\frac{\langle\chi\rangle^\ast}{f_\sigma}\right)^{\beta^\prime}\right\} \frac{g^\ast}{Z_{3B}^2}{\rm Tr}\left( \bar{\tilde{B}} \gamma_\mu \left[ \tilde{\rho}^\mu,  \tilde{B} \right] \right) \nonumber\\
& &{} - \left\{\tilde{g}_{V_1} + (g_{V_1} - \tilde{g}_{V_1})\left(\frac{\langle\chi\rangle^\ast}{f_\sigma}\right)^{\beta^\prime}\right\} \frac{g^\ast}{Z_{3B}^2}{\rm Tr}\left( \bar{\tilde{B}} \gamma_\mu \left\{ \tilde{\rho}^\mu,  \tilde{B} \right\} \right) .
\end{eqnarray}

From ~\eqref{eq:LagbsHLS} we have the terms contributing to the $g_A$ factor as
\begin{eqnarray}
{\cal L}_{bs{\rm HLS} }^{g_A} & = & {} - \tilde{g}_{A_1}{\rm Tr}\left( \bar{B} \gamma_\mu \gamma_5 \left\{ \mathcal{A}^\mu,  B \right\} \right) - \tilde{g}_{A_2}{\rm Tr}\left( \bar{B} \gamma_\mu \gamma_5 \left[ \mathcal{A}^\mu,  B \right] \right) \nonumber\\
& & {} - \Big[(g_{A_1} - \tilde{g}_{A_1}){\rm Tr}\left( \bar{B} \gamma_\mu \gamma_5 \left\{ \mathcal{A}^\mu,  B \right\} \right) + (g_{A_2} - \tilde{g}_{A_2}){\rm Tr}\left( \bar{B} \gamma_\mu \gamma_5 \left[ \mathcal{A}^\mu,  B \right] \right) \Big]\left(\frac{\langle\chi\rangle^\ast}{f_\sigma}\right)^{\beta^\prime} ,
\label{LBCT2}
\end{eqnarray}
\end{widetext}
Keeping only the nucleons, the Lagrangian is rewritten as
\begin{eqnarray}
{\cal L}_{bs{\rm HLS} }^{g_A} & = & {} - \tilde{g}_{A}\bar{N} \gamma_\mu \gamma_5 \mathcal{A}^\mu N \nonumber\\
& &{} - (g_{A} - \tilde{g}_{A})\left(\frac{\langle\chi\rangle^\ast}{f_\sigma}\right)^{\beta^\prime} \bar{N} \gamma_\mu \gamma_5 \mathcal{A}^\mu  N ,
\end{eqnarray}
where $\tilde{g}_A = \tilde{g}_{A_1} + \tilde{g}_{A_2}$. Therefore, the effective axial-vector coupling is expressed as
\begin{eqnarray}
g_A^{\ast} & = & {} \frac{1}{Z_{3B}^2}\left[\tilde{g}_{A} + (g_{A} - \tilde{g}_{A}) \left(\frac{\langle\chi\rangle^{\ast}}{f_\sigma}\right)^{\beta^\prime} \right].
\end{eqnarray}

After the derivation of the general expressions of the hadron properties in medium, we analyze these expressions with respect to the present status of the experimental measurement and other theoretical studies of medium modified hadron properties in the following section.



\section{Implication of the low energy constants}

\label{sec:impli}

The above derivation gives the medium modified hadron properties in the ``bare parameters" in  the most general effective Lagrangian.
As we noticed before, one of the disadvantageous of general effective Lagrangian is that there are so many
low energy constants that cannot be estimated at this moment that it is not feasible to make full numerical calculations. We will therefore rely on mean-field approximations and discuss the validity of the leading order scale symmetry. We can of course go beyond the mean field approximation and do more rigorous double-decimation RG analysis (such as $V_{lowk}RG$ applied in compact stars~\cite{PKLMR}).

Before going into details, we stress once more  that the estimation that follows of the low energy constants and the IDD of hadron properties is based on the ``bare" Lagrangian. If one wants to confront the experimental data by using the HLS$_\sigma$ and $bs$HLS, one has to sum over all the contributions for the present obtained results, the induced IDD as well as the higher order nuclear correlations~\cite{Paeng:2015noa,PKLMR}. In what follows, we will assume that the mean-field approximation corresponding to the single-decimation RG is applicable.

For later convenience, we define
\begin{eqnarray}
\Phi & = & {} \frac{\langle\chi\rangle^{\ast}}{f_\sigma} \lesssim 1.
\end{eqnarray}

\subsection{Low energy constants from the medium modified hadron spectrum}

Let us first consider the pion properties in medium. How the pion behaves in medium is an extremely subtle issue
because the pion is almost a perfect Nambu-Goldstone boson. In the chiral limit, its properties must be protected by chiral invariance. While the pion decay constant
will be affected by density, its mass will be protected by the symmetry and will not be affected by density.
There is an indication as to how the pion decay constant behaves in medium~\cite{Yamazaki:1996zb} and this
could be explained by a variety of models that have chiral symmetry in the chiral limit.  But how the pion
mass behaves is a different matter. It will depend on the models used with the chiral symmetry breaking incorporated.
Depending on the models, it could move upwards or downwards~\cite{Toki:1990fc,Hirenzaki:1991us,Drews:2016wpi,Drews:2016wpi}. Given the validity of the Gell-Mann,
Oakes, and Renner (GMOR) relation in medium, the quark condensate and pion decay constant are locked to each other, it is reasonable to regards the pion mass as a density independent quantity.
In our approach, this issue can be addressed only in the LOSS as was done in~\cite{Paeng:2015noa} with further corrections as an open problem.
Then, from the bare HLS$_\sigma$ Lagrangian, we get
\begin{eqnarray}
\frac{m_\pi^{\ast2}}{m_\pi^2} & = & \frac{\Phi^{1-\gamma_m}}{h_1+ (1 - h_1)\Phi^{\beta^{'}} } = 1 .\label{eq:pimpimr}
\end{eqnarray}
At this moment, there is no well established result on $\gamma_m$ of QCD at nonperturbative region. Here, we simply take $\gamma_m =1$. Then, the scaling \eqref{eq:pimpimr} indicates that $h_1 = 1$ and this choice is consistent with the requirement from the $\sigma \to \pi\pi$ decay in the matter free space~\cite{Crewther:2013vea}.

The experiment E325 performed at the KEK 12 GeV Proton Synchrotron measures the invariant mass spectra of $\rho,\omega, \phi  \to {\rm{ }}{{\rm{e}}^{\rm{ + }}}{{\rm{e}}^ - }$ and $ \phi  \to {\rm{ }}{{\rm{K}}^{\rm{ + }}}{{\rm{K}}^ - }$ decay modes simultaneously ~\cite{Naruki:2006zz,Sakuma:2006xc} found that the vector meson masses decrease with nuclear density. A similar conclusion was obtained in the upgraded CERES experiment at CERN ~\cite{Marin:2008aa}. Then, by using the HLS$_\sigma$ at tree level, we obtain
\begin{eqnarray}
\frac{m_{\rho,\phi}^{\ast 2}} {m_{\rho,\phi}^{ 2}}& = & \frac{\left[h_2+ (1 - h_2)\Phi^{\beta^{'}}\right]}{\left[ h_3+ (1 - h_3)\Phi^{\beta^{'}}\right]}\Phi^{2} < 1.
\end{eqnarray}
A simple choice of the parameters $h_2$ and $h_3$ satisfying the above constraint is $h_2 = h_3 = 1$. With this choice, one can predict that
\begin{eqnarray}
\frac{\Gamma^\ast\left( \phi \rightarrow e^+ e^- \right)}{\Gamma\left( \phi \rightarrow e^+ e^- \right)} & = &
\left\vert \frac{g_\phi^\ast}{m_\phi^{\ast2}}
\right\vert^2 \frac{m_\phi^{\ast}}{m_\phi}\left\vert \frac{g_\phi}{m_\phi^{2}}
\right\vert^{-2} = \Phi
\ ,
\label{phieewidth}
\end{eqnarray}
which indicates that the partial width decreases with density. Although this process has been measured at the KEK-PS E325 experiment~\cite{Sakuma:2008zz}, the density dependence cannot be well established due to the experimental uncertainty. So that, our this prediction cannot be excluded. A possible way to confirm the present result is to resort to the relative density dependence of $m_\rho^\ast$ and $\Gamma^{\ast}\left( \phi \rightarrow e^+ e^- \right)$ which is predicted the same here.

By matching the HLS to the QCD in the Wilsonian sense, it was found that near the chiral restoration scale, the vector manifestation of the Wigner realization of chiral symmetry characterized by $a \to 1$ is approached~\cite{Harada:2000kb}. If this condition from the basic QCD is accepted, one has
\begin{eqnarray}
\frac{a^{\ast}}{a} & = & \left[ h_2 + (1-h_2) \Phi^{\beta^\prime}\right] < 1,
\end{eqnarray}
at tree level where $h_1 = 1$ has been used. To satisfy this requirement near the chiral restoration scale, here regarded as density, one should have $h_2 < 1$. However, since our scaling relations as written are valid for the density $\lesssim n_{1/2} \simeq 2 n_0$, they cannot be relevant to the chiral restoration. So that, it is reasonable to take $a^\ast/a \simeq 1$ which agrees with the parameter choice $ h_2 =1$. With such a set of parameter choice, the density dependence of the vector meson mass is locked to that of pion decay constant.

The last particle we will consider in the mesonic sector is the dilaton. From Eq.~\eqref{eq:msigmas} we have
\begin{eqnarray}
\frac{m_\sigma^{\ast2} f_\sigma^{\ast2}}{m_\sigma^{2} f_\sigma^{2}} & = & \Phi^{4 + \beta^\prime},
\end{eqnarray}
which is a quantity independent of the low energy constants $h_4,h_5$ and $h_6$. Without the lose of generality, we take $h_4 = 1$. Therefore,
\begin{eqnarray}
\frac{m_\sigma^{\ast 2}}{m_\sigma^{2}} & = & \Phi^{2+\beta^\prime}, \quad
\frac{f_\sigma^{\ast}}{f_\sigma^{}} = \Phi.
\end{eqnarray}
Given the GMOR type relation $m_\sigma^2 f_\sigma^2 = 4 c \beta^\prime (4+ \beta^\prime)$ still survives in medium, an indirect consequence of the this parameter choice is that the parameter $h_5$ and/or $h_6$ are density dependent quantities which can be reduced from the scaling
\begin{eqnarray}
\frac{c^{\ast}}{c} & = & \Phi^{4+\beta^\prime},
\end{eqnarray}
which yields
\begin{eqnarray}
\frac{h_5^{\ast}}{h_5} & = & \Phi^{4+2\beta^\prime}, \quad \frac{h_6^{\ast}}{h_6} = \Phi^{4+\beta^\prime}.
\end{eqnarray}

Now, let's see the medium modified mass of baryon given by Eq.~(\ref{eq:massofb}). The ratio of $m_B^*$ and $m_B$ from the bare Lagrangian $bs$HLS is
\begin{eqnarray}
\frac{m_B^\ast}{m_B} & = & {} \frac{\left[\tilde{\mathring{m}}_B  + (\mathring{m}_B - \tilde{\mathring{m}}_B) \Phi^{ \beta^\prime} \right]}{\mathring{m}_B\left[g_1+ (1 - g_1)\Phi^{\beta^{\prime}}\right]}\Phi.
\label{eq:massratioofb}
\end{eqnarray}
So far, from the theoretical study, we know that baryon mass is reduced in the nuclear matter~\cite{Detar:1988kn,Glozman:2012fj,Paeng:2013xya,Ma:2013ooa,Ma:2013ela}. However, one cannot obtain any information from this reduction on the constant $\mathring{m}_B, \tilde{\mathring{m}}_B$ and $g_1$ since $\Phi < 1$, therefore, one can simply chose $g_1 = 1$ and $\mathring{m}_B = \tilde{\mathring{m}}_B$ and obtain
\begin{eqnarray}
\frac{m_B^\ast}{m_B} & = & {} \Phi,
\end{eqnarray}
which agrees with that obtained in the LOSS.

Next let us look at the medium modified nucleon-pion coupling $g_A^\ast$. From Eq.~\eqref{eq:SIGBB} we have
\begin{eqnarray}
g_A^{\ast} & = & {} \frac{1}{Z_{3B}^2}\left[\tilde{g}_{A} + (g_{A} - \tilde{g}_{A}) \Phi^{\beta^\prime} \right].
\end{eqnarray}
If we take the LOSS, we have
\begin{eqnarray}
g_1 & = & 1, \quad \tilde{g}_{A} = g_{A}.
\end{eqnarray}
As a result, $g_A^\ast$ is a density invariant quantity. In~\cite{LMRLetter}, this prediction is shown to present, backed by a quantum {\it ab initio} Monte Carlo calculation in nuclei by Pastore et al~\cite{pastore},   an extremely simple solution to the long-standing ``$g_A$ quenching problem" in nuclear Gamow-Teller transitions, i.e., giant Gamow-Teller resonances, double beta decays etc. This is one of the two confrontations of the LOSS in nuclear physics, the other being in the EoS of massive compact stars~\cite{PKLMR}.

Next we consider the nucleon-sigma coupling in medium. For this purpose, we regard the nucleon in medium as a nearly on-mass-shell particle. Then, from Eq.~(\ref{eq:SIGBB}), considering the Dirac equation and the parameter choice discussed above, we have
\begin{eqnarray}
{g^*}_{\sigma\bar{B} B} & \simeq & m_B^\ast\frac{1}{f_\sigma^\ast} = \frac{m_B}{f_\sigma},
\end{eqnarray}
which is a scale independent quantity  determined from the Goldberger-Trieman type relation in matter free space.

\subsection{Effective hadron coupling}

After determining the low energy constants, we now predict the medium modified coupling constants. For the strong interaction in the mesonic sector, we have
\begin{eqnarray}
{\cal L}_{\sigma\pi\pi } & = & 2 \frac{ 1}{f_\sigma^\ast}\tilde{\sigma} {\rm Tr}\left( \partial_\mu \tilde{\Pi} \partial^\mu \tilde{\Pi} \right) -\frac{2}{ f_\sigma^\ast} \tilde{\sigma} {\rm Tr}\left(\mathcal{M}^{\ast}\tilde{\Pi}^2 \right), \\
{\cal L}_{\sigma\sigma\sigma} & = & \frac{1}{f_\sigma^\ast}\tilde{\sigma}\partial_\mu \tilde{\sigma} \partial^\mu \tilde{\sigma} \nonumber\\
& &{} + \frac{2}{3} \Big[ c^{\ast}(4+\beta^\prime)\left[16 - (4+\beta^\prime)^2\right] \nonumber\\
& &\qquad \; {} + f_{\pi}^{2} (2m_K^{ 2} + m_\pi^{ 2})\Phi^{2} \Big]\frac{1}{f_\sigma^{\ast 3}} \tilde{\sigma}^3 , \nonumber\\
{\cal L}_{\rho\pi\pi} & = & i a g{\rm Tr}[\tilde{\rho}_\mu \left[\partial^\mu \tilde{\pi}, \tilde{\pi}\right]],\nonumber\\
{\cal L}_{\sigma\rho\rho} & = &{} 2a g^{ 2}f_\pi^{\ast 2} \frac{1}{f_\sigma^\ast}\tilde{\sigma}{\rm Tr}[\tilde{\rho}_\mu \tilde{\rho}^\mu] .
\end{eqnarray}
These expressions explicitly show that, $\rho$-$\pi$-$\pi$ coupling is not touched by matter effect, but both $\sigma$-$\pi$-$\pi$ and $\sigma$-$\rho$-$\rho$ coupling scale as $\Phi$. However, the scaling of $\sigma$-$\sigma$-$\sigma$ coupling is complicated. In chiral limit, it reduces to
\begin{eqnarray}
{\cal L}_{\sigma\sigma\sigma} & = & \frac{1}{f_\sigma^\ast}\tilde{\sigma}\partial_\mu \tilde{\sigma} \partial^\mu \tilde{\sigma} \nonumber\\
& &{} + \frac{2}{3}\frac{c^{\ast}}{f_\sigma^{\ast 3}}(4+\beta^\prime)\left[16 - (4+\beta^\prime)^2\right] \tilde{\sigma}^3 ,
\end{eqnarray}
which means that, for the low energy process in which the off-shell effect of sigma meson is small, the $\sigma$-$\sigma$-$\sigma$ coupling scales as $\Phi^{\beta^\prime + 1}$.

For the effective electroweak coupling, we have
\begin{eqnarray}
g_{\rho\gamma}^\ast & = & g_{\rho\gamma} \Phi^{2} = \frac{m_\rho^{\ast 2}}{g^\ast},\nonumber\\
g_{\gamma\pi\pi}^{\ast} & = & e \left(1 - \frac{a}{2} \right) .	
\end{eqnarray}
The last equation indicates that the vector meson dominance is still valid in medium at low density in the LOSS at tree level.

We finally turn to the interaction in the baryonic sector. With the parameter choice discussed above, we have
\begin{eqnarray}
{\cal L}_{\sigma\bar{B} B} & = & - \mathring{m}_B \frac{1}{f_\sigma}\tilde{\sigma}{\rm Tr}\left( \bar{\tilde{B}} \tilde{B}\right), \nonumber\\
{\cal L}_{\pi\bar{B} B} & = &{} - \frac{1}{f_\pi^\ast}g_{A_1} {\rm Tr}\left( \bar{\tilde{B}} \gamma_\mu \gamma_5 \left\{ \partial^\mu\tilde{\pi}, \tilde{B}\right\} \right) \nonumber\\
& &{} - \frac{1}{f_\pi^\ast} g_{A_2} {\rm Tr}\left( \bar{\tilde{B}} \gamma_\mu \gamma_5 \left[ \partial^\mu\tilde{\pi}, \tilde{B}\right] \right) ,\nonumber\\
{\cal L}_{V\bar{B} B} & = & \left(1 - g_{V_2}\right) g {\rm Tr}\left( \bar{\tilde{B}} \gamma_\mu \left[ \tilde{\rho}^\mu,  \tilde{B} \right] \right) \nonumber\\
& &{} - g_{V_1} g {\rm Tr}\left( \bar{\tilde{B}} \gamma_\mu \left\{ \tilde{\rho}^\mu,  \tilde{B} \right\} \right) .
\end{eqnarray}
These expressions show that the interaction strengths of $\sigma$-$B$-$B$ and $\rho$-$B$-$B$ are not affected by medium effect but that of $\pi$-$B$-$B$ scales as $\Phi^{-1}$. And, with this parameter choice discussed above from the Ward identity between the axial-vector and psuedo-scalar currents, we obtain
\begin{eqnarray}
g_{\pi BB}^\ast & = &{} \Phi^{3-\gamma_m}\frac{1}{Z_{3B}^2 Z_{3\pi}}\frac{m_B^\ast}{f_\pi^\ast}g_A^\ast = \frac{m_B}{f_\pi}g_A,
\end{eqnarray}
which is a density invariant quantity. This relation indicates that, with the present parameter choice, the Goldberger-Trieman relation is protected in medium.

\section{Summary and discussion}

\label{sec:dis}

In this work, we analyzed the low energy constants appearing in the chiral-scale effective theory  in the mean-field approximation which we believe is applicable for density $n\sim 2n_0$. We find that to be consistent with the experimental measurement performed so far and the requirement from QCD, the hadron properties have the following scaling
\begin{eqnarray}
\frac{m_\pi^{\ast}}{m_\pi} & = & \Phi^0, \quad \frac{m_\sigma^{\ast}}{m_\sigma} = \Phi^{\beta^\prime/2 +1 } ,\quad
\frac{m_B^{\ast}}{m_B} = \frac{m_\rho^{\ast}}{m_\rho} = \frac{f_{\pi}^{\ast}}{f_{\pi}} = \Phi \nonumber\\
\frac{h_5^{\ast}}{h_5} & = & \Phi^{4+2\beta^\prime}, \quad
\frac{h_6^{\ast}}{h_6} = \Phi^{4+\beta^\prime}, \nonumber\\
\frac{g_{\sigma\pi\pi}^{\ast}}{g_{\sigma\pi\pi}} & = & \Phi^{-1}, \quad
\frac{g_{\sigma\rho\rho}^{\ast}}{g_{\sigma\rho\rho}} = \Phi, \quad
\frac{g_{\sigma\sigma\sigma}^{\ast}}{g_{\sigma\sigma\sigma}} = \Phi^{\beta^\prime+1}, \nonumber\\
\frac{g_{\sigma BB}^{\ast}}{g_{\sigma BB}} & = & \frac{g_{\pi BB}^{\ast}}{g_{\pi BB}}= \frac{g_{\rho BB}^{\ast}}{g_{\rho BB}} = \frac{g_{\rho\pi\pi}^{\ast}}{g_{\rho\pi\pi}}  = \Phi^{0}.
\end{eqnarray}
Note that, we considered the $SU(3)$ flavor symmetry for pseudoscalar meson and baryons and $U(3)$ hidden local symmetry form vector mesons in the present work, so that scaling properties for other hadrons can be obtained by flavor rotation. Some of these relations, such as $m_\rho^\ast/m_\rho, m_B^\ast/m_B$, have been given in Ref.~\cite{PKLMR} based on the LOSS. And, when we take $\beta^\prime \ll 1 $ which is required by LOSS, upto the leading order of $\beta^\prime$ expansion, the scaling of $m_\sigma^\ast/m_\sigma$ obtained here is the same as that gotten in Ref.~\cite{PKLMR}. It would be interesting to look at nuclear matter properties such as the binding energy and equilibrium density with this scaling of sigma mass which scales as $\beta^\prime/2 + 1$ and omega mass which scales as $1$.

We want to say that, the reduction to the LOSS from the general scale-chiral effective theory and the scaling behaviours obtained here is valid, probably only up to the density $\sim n_{1/2} \simeq 2n_0$ which is determined from the skyrmion crystal approach to nuclear matter~\cite{Ma:2016gdd}. After $n_{1/2}$, there is a drastic change due to topology change, so one needs to be cautious on this formula as one reaches near the flash point of chiral restoration.

As stated, the IDD we are talking about here is the property inherited from QCD in the sense of Wilsonian matching between the effective theory and QCD such that the bare parameters in the effective theory become background, here density, dependent. When one wants to confront the experimental measurement to the theoretical quantities, such as the medium modified mass, medium modified decay constant, one should notice that measurements are correlation functions and the interpretation of the density dependence of mass and coupling constants are dependent on the specific models~\cite{Brown:2005ka,Brown:2005kb,Brown:2009az}.  Therefore it is meaningless -- and sometimes totally erroneous --  to interpret certain theoretical approaches, for example Refs.~\cite{Kolomeitsev:2002gc,Drews:2016wpi,Goda:2013npa},  in terms of the BR scaling defined in the given ``bare" effective  Lagrangian. The bare IDD, the induced density effects from  short-range multibody forces as well as all higher order nuclear correlations should be taken into account to make a proper comparison.

Since the LOSS is compatible with the present constraint from the experimental measurement, this gives a dramatic simplification of the higher scale-chiral order Lagrangian, such as those given in Ref.~\cite{Li:2016uzn}. In addition, from the general construction of Ref.~\cite{Li:2016uzn}, one can go beyond the LOSS and  take into account order by order with the scale symmetry breaking together with chiral symmetry breaking.

A compelling confirmation of the validity of the LOSS is given in~\cite{LMRLetter}. There the solution is given to  the long-standing -- some four-decade old -- mystery of ``$g_A$ quenching" in nuclei~\cite{Wilkinson:1973zz,Wilkinson:1973mgv,Wilkinson:1974huj,Buck:1975ae}. It also provides the most pristine signal for soft-pion exchanges in nuclei. Together with the ``chiral filter mechanism"~\cite{KDR}  given a quantitative support, it is perhaps the most compelling evidence of Brown-Rho scaling in nuclear matter.  The relations derived here will serve not only to unravel the properties of hadrons under extreme conditions with the possibility of making systematic higher-order scale-chiral corrections, but also expose the hidden  symmetries of QCD, scale symmetry and  local flavor symmetry, a fundamental issue in particle and nuclear physics.

\subsection*{Acknowlegments}

We would like to thank Mannque Rho for his valuable discussions and comments. Y.~L. Ma was supported in part by National Science
Foundation of China (NSFC) under Grant No. 11475071, 11547308 and the Seeds Funding of Jilin
University.


\end{document}